\documentclass[aps,a4paper,twocolumn,floatfix,showpacs,
groupedaddress]{revtex4}

\usepackage{bm,amsmath,amssymb,amsfonts}

\usepackage[dvips]{graphicx}

\usepackage{color}
\definecolor{dblue}{rgb}{0.0,0.0,0.7}

\begin{document}

\title{\textcolor{dblue}{Staggered and extreme localization of electron states in fractal space}}

\author{Biplab Pal}

\email[\textbf{E-mail:} biplabpal@klyuniv.ac.in]{}

\author{Arunava Chakrabarti}

\affiliation{Department of Physics, University of Kalyani, Kalyani,
West Bengal - 741 235, India}

\begin{abstract}
We present exact analytical results revealing the existence of a countable 
infinity of unusual single particle states, which are localized with a multitude of 
localization lengths in a Vicsek fractal network with diamond shaped loops 
as the `unit cells'. The family of localized states form clusters of 
increasing size, much in the sense of Aharonov-Bohm cages [J. Vidal {\it et al.}, 
Phys. Rev. Lett. \textbf{81}, 5888 (1998)], but now without a magnetic field.
The length scale at which the localization effect for each of these states 
sets in can be uniquely predicted  following a well defined prescription 
developed within the framework of real space renormalization group. 
The scheme allows an exact evaluation of the energy eigenvalue for every such 
state which is ensured to remain in the spectrum of the system even in the 
thermodynamic limit. In addition, we discuss the existence of a perfectly 
conducting state at the band center of this geometry and the 
influence of a uniform magnetic field threading each elementary plaquette of 
the lattice on its spectral properties. Of particular interest is the case of 
extreme localization of single particle states when the magnetic flux equals 
half the fundamental flux quantum. 
\end{abstract}

\maketitle

\section{Introduction}
Interplay of lattice topology and quantum interference effects is known to give 
rise to exotic electronic spectrum in solid systems that has been studied in 
details over several decades by now. The subject is still being pursued with vigor 
and with an aim to achieve comprehensive control over coherent transport in low 
dimensions. 

In quantum interference and related transport mechanism a pivotal role is played  
by the localization of electronic eigenstates in presence of disorder. Such 
localization, known as the Anderson localization~\cite{anderson} upholds a central 
result that, in one dimension with arbitrary disorder, all the single particle 
states will be exponentially localized, and the same was shown to ring true in two 
dimensions as well~\cite{abrahams}. Since then, extensive research has been 
undertaken to understand the fundamentals of localization effects, studies ranging 
from electronic states in random lattice models~\cite{angus}-\cite{zilly}, to the 
Anderson localization of light~\cite{jovic,bliokh}, spin freezing in one 
dimensional semiconductors~\cite{echeverria}, and localization in optical 
lattices~\cite{sankar1}, to name a few. Matter waves can also be localized in 
deterministic potentials sharing certain features of random 
disorder~\cite{aubry,sankar2}. Recent experiments reveal the Anderson localization 
of non-interacting Bose-Einstein condensates in one dimensional matter waveguides, 
where the random potential has been generated by laser speckles~\cite{billy}. 
Similar experiments have also been reported to study the Anderson localization 
in optical lattices~\cite{edwards,roati}, and in the cases of 
microwaves~\cite{shi}  and of classical waves in weakly disordered one dimensional 
stack of meta-materials~\cite{asatryan}.

Variations of the classic Anderson localization are also well known by now.  
Isolated de-localized (extended) single particle states exist, even in a 
disordered one dimensional chain of atomic potentials, resulting out of a kind of 
spatial correlation~\cite{dunlap, moura1}, in 1-d quasiperiodic 
chains~\cite{arunava1}-\cite{macia3}, or, in certain kinds of deterministic 
fractal geometries~\cite{arunava3}-\cite{schwalm1}. Crossover from an insulating 
to a metallic spectral behavior in correlated disordered two-legged ladder 
networks have also been reported recently~\cite{sil1,moura2}.

A curious point, apparently gone unnoticed or un-appreciated so far is that, while 
a precise determination of the eigenvalues corresponding to the extended single 
particle states is possible in the above cases of correlated and deterministic 
disorder, the task seems to be practically impossible when it comes to an exact 
evaluation of eigenvalues of the localized states in a random or even a 
deterministically disordered system in the {\it thermodynamic limit}. 
It should be appreciated that, though a direct diagonalization of the Hamiltonian 
for a finite size of the system yields eigenvalues of the localized states (for 
a disordered or a deterministically disordered system), there is no apriory reason 
to assume that these eigenvalues remain in the spectrum when the system grows 
in size, and tends to infinity. In fact, for a deterministic fractal geometry 
that offers a singular continuous spectrum, it is almost impossible to hit the 
exact eigenvalues corresponding to the states that will finally be 
localized on an infinite lattice. To the best of our knowledge, this issue 
remains unaddressed so far in the literature. 

Can one really identify the localized states and extract the corresponding 
eigenvalues for a deterministically disordered system?
In the present communication we address ourselves this question, and  
take up the task of critically examining the spectral properties  of a 
Vicsek fractal network~\cite{vicsek} consisting of {\it diamond} shaped loops 
within a tight binding formalism. While looking for the localized state 
eigenvalues and the nature of localization are indeed the major factors driving 
this work, other interests in such a study are related to the general spectral 
character and magneto-transport in such systems. The motivation behind the latter 
part of the work may be summarised as follows. A {\it diamond-Vicsek} network (see 
Fig.~\ref{system}) provides an interesting geometry in which the `open' character 
of a typical Vicsek pattern is preserved along with the presence of closed loops 
in shorter scales of length. This is in marked contrast to the much studied 
Sierpinski gasket~\cite{rammal,domany}, which is a closed structure, or to the 
other open tree fractals~\cite{tree} or even an alternative version of the Vicsek 
fractal without any local closed loops~\cite{jayanthi}. The presence of these 
loops effectively generates a longer ranged interaction between the atomic sites 
occupying the various vertices, and its effect on the electron localization or 
de-localization is worth studying. 

Secondly, linear arrays of diamond networks have already drawn considerable 
attention in recent years in the context of charge and spin transport 
properties~\cite{sil2}-\cite{aharony2}, being shown to behave as a {\it flux 
controlled} $n$- or $p$-type semiconductor~\cite{sil2}, or as a prospective 
candidate of an elegant spin filter~\cite{aharony1,aharony2}. The influence of a 
topological variation in the arrangement of such loops on the spectral properties 
of the system is thus worth investigating, both from the standpoint of fundamental 
physics, and from the perspective of device technology. We choose 
such a deterministic geometry to make an analytical attack on the system possible.

We find extremely interesting results in the context of localization of electronic 
states. In the absence of any external magnetic field, a countable infinity of 
localized states can be precisely detected with a multitude of localization 
lengths. One can work out an exact mathematical prescription to specify the length 
scale at which the onset of localization takes place. The localization can in 
principle, be {\it delayed} (staggered) in position space and  the corresponding 
energy eigenvalues can be exactly evaluated following the same prescription based 
on a real space renormalization group (RSRG) method. 
In addition, it is shown that for a given set of parameters, the center of the 
spectrum corresponds to a perfectly extended eigenstate, with the parameters of 
the Hamiltonian exhibiting a fixed point behavior. Switching on a magnetic field 
opens up  gaps in the spectrum in general, and even leads to an extreme 
localization of all the single particle states in the sense of formation of the 
Aharonov-Bohm cages~\cite{vidal}. 

In what follows we describe the results. In section II, the model and the 
mathematical method of handling the problem are presented. Section III and IV 
include the results and their analyses related to the 
spectral properties without and with the magnetic field respectively. 
The two terminal transport study is carried out in section V and in 
section VI we draw our conclusions. 
\section{The system and the mathematical formulation}
\subsection{The Hamiltonian}
\begin{figure*}[ht]
\includegraphics[clip,width=16cm]{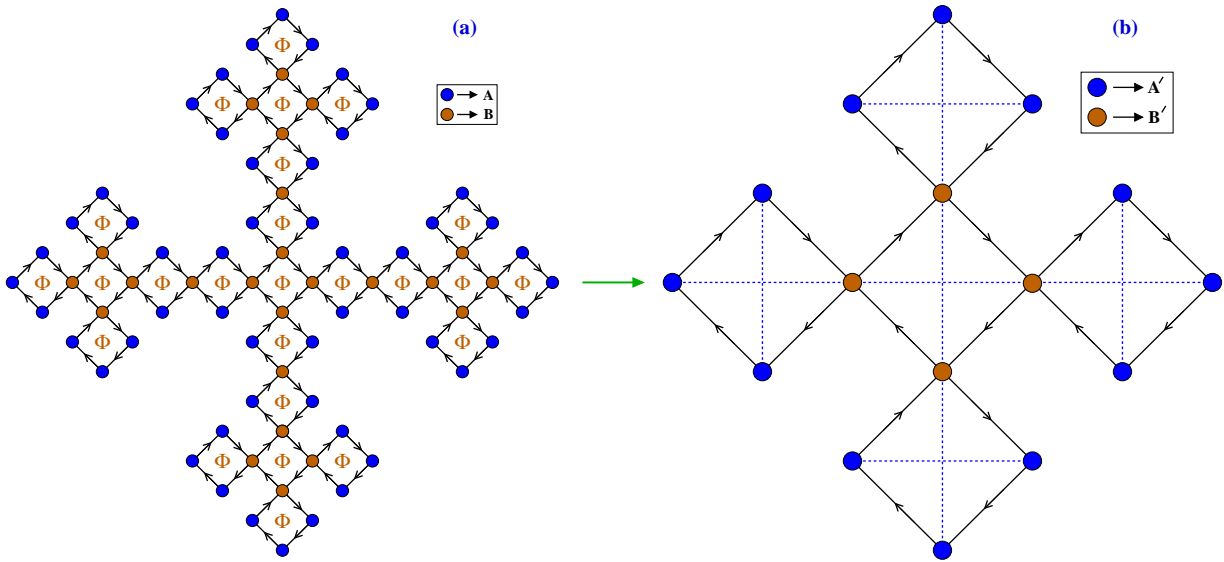}
\caption{ (Color online) (a) Schematic view of the second generation of an 
infinite diamond-Vicsek network with each diamond plaquette threaded by a uniform 
magnetic flux $\Phi$. The edge sites are named `$A$' (blue circles) and the 
bulk sites are marked as `$B$' (orange circles), and the arrow indicates the 
direction of the forward hopping. 
We exclude any second neighbor hopping at the beginning.
(b) Renormalized version of (a) with the dotted 
lines indicating the diagonal hopping which is generated due to renormalization.}
\label{system}
\end{figure*}

We refer to Fig.~\ref{system}(a) 
which illustrates the second generation of a Vicsek 
geometry with diamond shaped loops. As, at one stage we shall be considering the 
effect of a magnetic field on the spectral properties, we show in the figure the 
flux distribution. Each plaquette is threaded by a uniform 
magnetic flux $\Phi$. The atomic sites are assigned different status depending on 
their positions and neighborhood in the lattice, namely, the sites in the bulk are 
marked as `$B$', while `$A$' refers to the sites sitting at the edges. The 
magnetic field breaks the time reversal symmetry along the edges of every diamond. 
The Hamiltonian, in the tight binding formalism is written as,
\begin{equation}
\mbox{\boldmath $H$}=\sum_{i}\epsilon_{i}|i\rangle\langle{i}|
+\sum_{\langle ij \rangle} \:[\;t_{ij} e^{i \theta_{ij}}|i \rangle 
\langle{j}| + t_{ji} e^{-i \theta_{ij}}|j \rangle \langle{i}|\;]
\label{Hamiltonian}
\end{equation}
where, $\epsilon_{i}$ is the on-site potential at the $i$-th site, and has a value 
$\epsilon_{A}$ or $\epsilon_{B}$ depending on whether its an `edge' site or a 
`bulk' one. The uniform nearest neighbor hopping amplitude is $t_{ij}=t$ along the 
edges, and $t_{ij} = \lambda$ when $i$ and $j$ refer to the opposite vertices, 
connected by a diagonal. Thus we keep the provision of including hopping beyond 
the nearest neighbors.  $\theta_{ij}$ is the Peierls' phase~\cite{peierls} given 
by, $\theta_{ij} = 2 \pi \Phi/4 \Phi_0$ for hopping along an edge. $\Phi_0 = hc/e$ 
is the fundamental flux quantum. From symmetry considerations, $\theta_{ij} = 0$ 
when the electron hops across a diagonal, that is, when $t_{ij} = \lambda$.
\subsection{The RSRG scheme}
An elegant way of handling such self-similar systems is to use the real space 
renormalization group (RSRG) method~\cite{southern} where one can decimate out a 
subset of atomic sites from the original lattice to get a scaled version of it 
(Fig.~\ref{system}(b)). This is easily done by writing down in details the set of 
difference equations, 
\begin{equation}
(E-\epsilon_{i})\:\psi_{i}=\sum_{j}\;t_{ij} e^{i \theta_{ij}}\:\psi_{j}
\label{difference}
\end{equation}
where, $\psi_i$ denotes the amplitude of the wave function at the $i$-th site, and 
$\theta_{ij}$ is the Peierls' phase associated with the hopping matrix element 
connecting the $i$-th and the $j$-th sites. We begin with nearest neighbor 
hopping only (that is, we set $\lambda = 0$ at the beginning).
However, such a decimation automatically generates the second neighbor hopping 
across the  diagonals of an inflated diamond as shown by the dotted line in 
Fig.~\ref{system}(b). The range of interactions of course does not increase beyond 
this on further renormalization. It is therefore advisable to retain $\lambda$ in 
the Hamiltonian from the very beginning. One can easily compare the results 
obtained by switching $\lambda$ {\it on} or {\it off}. The recursion relations of 
the on-site potentials and the hopping integrals are provided below. 
\begin{eqnarray}
\epsilon_{A}^{\prime} &=& \epsilon_{A}+
[pt_{f}+p^{\ast}t_{b} + \alpha \lambda_{1}] \nonumber \\  
\epsilon_{B}^{\prime} &=& \epsilon_{B}+
2\,[pt_{f}+p^{\ast}t_{b} + \alpha \lambda_{1}] \nonumber \\
t_{f}^{\prime} &=& \beta \lambda_{1};\ 
t_{b}^{\prime} = \beta^{\ast} \lambda_{1};\ 
\lambda^{\prime} = \gamma \lambda_{1} 
\label{recursion}
\end{eqnarray}

where, 
$\alpha =[(E-\bar{\epsilon}_{B})\lambda_{1}]/\xi_{3}$; 
$\beta = [(E-\bar{\epsilon}_{B})\bar{t}_{f}+
\lambda_{2}\bar{t}_{b}]/\xi_{3}$; 
$\beta^{\ast} = [(E-\bar{\epsilon}_{B})\bar{t}_{b}+
\lambda_{2}\bar{t}_{f}]/\xi_{3}$; 
$\gamma = \lambda_{1}\lambda_{2}/\xi_{3}$; 
$\xi_{3} = (E-\bar{\epsilon}_{B})^{2} - \lambda_{2}^{2}$,
\vskip .1in
with, 
$\bar{\epsilon}_{B} = \tilde{\epsilon}_{B}+
w^{\ast}t_{f}+wt_{b}$; 
$\bar{t}_{f} = ut_{f}+vt_{b}$; 
$\bar{t}_{b} = vt_{f}+ut_{b}$;
$\lambda_{2} = \lambda +wt_{f}+w^{\ast}t_{b}$.
\vskip .1in
Here, 
$u = [(E-\tilde{\epsilon}_{B})\lambda_{1}]/\xi_{2}$; 
$v = \lambda \lambda_{1}/\xi_{2}$; 
$w = [(E-\tilde{\epsilon}_{B})t_{f}+\lambda t_{b}]/\xi_{2}$;
$w^{\ast} = [(E-\tilde{\epsilon}_{B})t_{b}+
\lambda t_{f}]/\xi_{2}$;
$\xi_{2} = (E-\tilde{\epsilon}_{B})^{2} - \lambda^{2}$,
\vskip .1in
with, 
$\tilde{\epsilon}_{B} = \epsilon_{B}+
pt_{f}+p^{\ast}t_{b}$;
$\lambda_{1} = \lambda +
p^{\ast}t_{f}+pt_{b}$ \\
and 
$p = [(E-\epsilon_{A})t_{b}+
\lambda t_{f}]/\xi_{1}$;
$p^{\ast} = [(E-\epsilon_{A})t_{f}+
\lambda t_{b}]/\xi_{1}$;
$\xi_{1} = (E-\epsilon_{A})^{2} - \lambda^{2}$.\\

In the above expression, $t_{f}= t_{b}^{\ast}= te^{i \theta}$, 
where $\theta = 2 \pi \Phi/4 \Phi_0$ is the constant Peierls' phase. 
The above recursion relations are then used to obtain information about the local 
density of states (LDOS) at specific sites of the system, and the character of the 
single particle states, as discussed below.
\section{Spectral properties with zero magnetic field}
\subsection{Local density of states in zero magnetic field and with \boldmath $\lambda = 0$}
\begin{figure}[ht]
\includegraphics[clip,width=10cm,angle=-90]{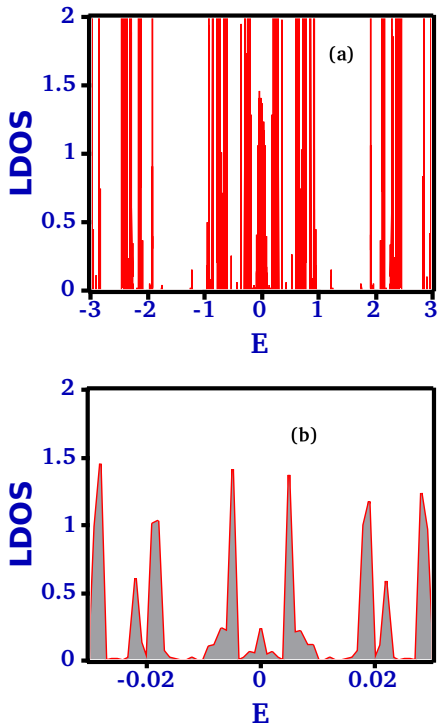}
\caption{ (Color online) (a) LDOS-$E$ plot at the bulk sites ($B$-type) of an 
infinite diamond-Vicsek network in absence of the magnetic field ($\Phi=0$). The 
other parameters are $\epsilon_{A}=\epsilon_{B}=0$, $t=1$ and $\lambda=0$. (b) A 
highlighted version of (a) around the center $E=0$.}
\label{ldos1}
\end{figure}

Using the standard decimation procedure~\cite{southern}, the LDOS at the edge 
($A$) and the bulk ($B$) sites can easily be obtained through the local Green's 
functions. For simplicity we present in Fig.~\ref{ldos1}(a) the LDOS at a $B$-site 
only, given by,  
\begin{equation}
\rho^{(B)}(E)= \lim_{\eta \rightarrow 0}\,\left[-\dfrac{1}{\pi}\,
\text{Im}\,\{G^{(B)}(E+i\eta)\}\right]
\label{dos}
\end{equation}
where, $G^{(B)}(E + i\eta) = (E + i\eta - \epsilon_B^*)^{-1}$, 
$\epsilon_B^*$ being the fixed point value of the relevant on-site 
potential at the $B$-site, obtained by iterating Eq.~\eqref{recursion}.
We have set $\epsilon_A = \epsilon_B =0$, $t = 1$ and $\lambda = 0$, and there is 
no magnetic field (i.e., $\Phi=0$). The LDOS shows a dense packing of eigenstates 
over a finite range of energy centered at $E = 0$. We have minutely examined this 
continuum by fine scanning an energy interval around $E = 0$, and show it in 
Fig.~\ref{ldos1}(b).
The continuum seems to persist. In the neighborhood of the 
band-center and within the apparent `continuum', the hopping integral  
remains non-zero over a substantial number of RSRG iteration steps. The number of 
such steps $n$ depends on the chosen energy and indicates that 
the corresponding eigenfunction is either of an {\it extended} character, or, at 
least, has very large localization length. This aspect will be further discussed 
in the following subsection in relation to the so called staggered localization 
effect. 

A particularly interesting state is the band center, viz., $E = 0$, where the 
entire parameter space $\{\epsilon_A, \epsilon_B, t, \lambda\}$ exhibits a one 
cycle {\it fixed point} behavior, viz., $\{\epsilon_A(n+1), \epsilon_B(n+1), 
t(n+1), \lambda(n+1)\}$ = $\{\epsilon_A(n), \epsilon_B(n), t(n), \lambda(n)\}$ for 
$n \ge 1$. $n$ stands for the RSRG iteration step. 
$\lambda$ at this special energy remains zero throughout the iteration.
We conclude that the eigenstate at the band-center is definitely {\it 
extended}, but is of a non-Bloch character. The general behavior of the hopping 
integrals under successive RSRG iterations is suggestive of the fact that this 
central extended eigenstate might be flanked on either side by a countable 
infinity of eigenstates which belong either to the extended category, or have very 
large localization lengths.
\subsection{Exact construction of eigenstates}
The inherent self-similarity of the deterministic fractals allows for the 
construction of exact distribution of amplitudes 
of the eigenstates, by suitably exploiting  
Eq.~\eqref{difference}. Previous attempts in this regard have 
unfolded extended non-Bloch states ({\it atypically extended} states) in the cases 
of an open loop Vicsek fractal~\cite{bibhas} or a closed loop diamond hierarchical 
geometry~\cite{anirban}. The present lattice 
offers a richer spectrum, allowing one to explicitly construct localized states 
extending over clusters of lattice points of various 
sizes on the parent lattice. The planar extent of 
such clusters depends on the eigenvalue corresponding to the localized state, and 
can be small or enormous.  

To elaborate, let us  consider the solutions of the equation, 
\begin{equation}
E = \epsilon_B(n) - 2 \lambda(n)
\label{roots}
\end{equation}
where, $n$ refers to the stage of renormalization. This is in general, a 
polynomial equation in $E$. The zero's of the polynomial will be eigenvalues of 
the infinite system if, and only if, with them one can satisfy 
Eq.~\eqref{difference} locally at every vertex of the lattice, even when the 
lattice grows infinitely large. This task can be accomplished by trying to 
draw a non-trivial distribution of amplitudes for an energy obtained from 
Eq.~\eqref{roots} on the undecimated vertices of an $n$-step renormalized lattice, 
and then trying to figure out the amplitude distribution on the original lattice 
at the bare length scale. Let us discuss two specific cases at first.

{\it Case I:} We begin with the un-renormalized lattice. Now  $n = 0$, and with 
$\epsilon_A = \epsilon_B =0$ and  $\lambda = 0$ the solution of 
Eq.~\eqref{roots} is $E = 0$. One can construct an eigenfunction for $E = 0$ 
with amplitudes equal to $\pm 1$ distributed alternately at the `$B$' sites along 
the major $X$- and $Y$-axes. The difference equation, viz., Eq.~\eqref{difference} 
can then easily be satisfied for all other intermediate vertices using the values 
$0$ or $\pm 1$.

\begin{figure*}[ht]
\includegraphics[clip,width=16cm]{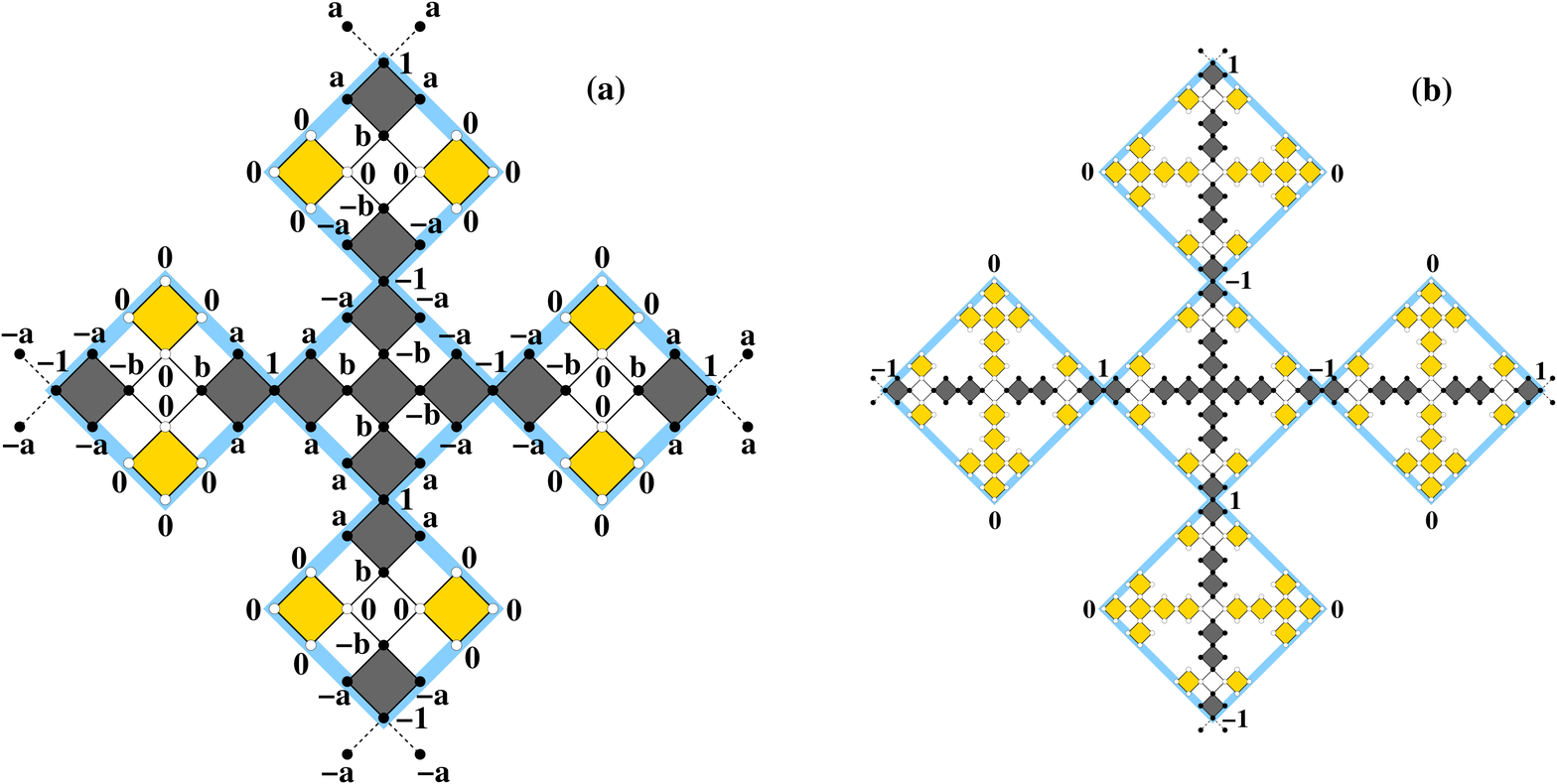}
\caption{ (Color online) (a) Distribution of amplitudes of the wave 
functions at $E=\sqrt{6}$ (obtained by solving Eq.~\eqref{roots} for $n=1$) 
on a $2$nd generation lattice. The dark shaded plaquettes (grey ones) embrace 
atomic sites with non-zero amplitudes (black solid circles) and the light 
shaded plaquettes (yellow ones) are surrounded by atomic sites with zero 
amplitudes (white empty circles). The initial parameters are 
chosen as $\Phi=0$, $\epsilon_{A}=\epsilon_{B}=0$, $t=1$ and $\lambda=0$. 
In the figure, $a=\sqrt{6}/4$ and $b=1/2$.
(b)~Distribution of amplitudes of the wave functions at $E=\sqrt{6}$ on a 
$3$rd generation lattice, the other parameters and symbols are same as (a). 
In the figures (a) and (b), the highlighted thick (blue) lines represent 
one-step and two-step renormalized lattice respectively.}
\label{ampdistribution1}
\end{figure*}

{\it Case II:} The above idea can indeed be extended to higher values of $n$, as 
we demonstrate in Fig.~\ref{ampdistribution1}(a)  
for $n = 1$, and discuss below. 
Let us extract the roots of the 
Eq.~\eqref{roots} for $n = 1$. The roots are 
$E = 0$ and $\pm \sqrt{6}$ for $\epsilon_{A}=\epsilon_{B}=0$, $t=1$ and 
$\lambda=0$ initially. We explain the construction 
of amplitudes for $E=\sqrt{6}$. 
The trick in this case is to place the values $\pm 1$ periodically along the major 
$X$-and $Y$-axes, and to assign an amplitude equal to {\it zero} at every `edge' 
($A$-type site) on a {\it one step renormalized} lattice.
The amplitudes at the intermediate sites of the original lattice are 
then systematically evaluated using Eq.~\eqref{difference}. 
We show it in Fig.~\ref{ampdistribution1}(a) on a 
small portion of an infinite lattice. 
The bigger square boxes with thick, highlighted edges represent the one step 
renormalized lattice. The construction depicted on a smaller 
scale can be extended to see that the distribution holds 
on a lattice of larger spatial extent (Fig.~\ref{ampdistribution1}(b)). In fact it
holds even on a lattice of an arbitrarily large size, 
where the {\it end} sites are not actually visible. As, according to our earlier 
argument, we are able to satisfy Eq.~\eqref{difference} locally at every vertex 
while drawing this distribution, $E=\sqrt{6}$ is definitely an eigenvalue of the 
infinite system, a fact that has been cross-checked by evaluation the LDOS at the 
$A$- and the $B$-sites at this special energy. We get a stable, finite value of 
the LDOS which supports our argument above.

Looking back at Fig.~\ref{ampdistribution1}, the solid 
black dots represent non-zero values of the amplitude while an open circle 
represents an amplitude equal to zero. Non-zero amplitudes, represented by solid 
circles, have the values equal to $\pm 1$, $\pm \sqrt{6}/4$ (depicted 
by the letter $a$) and $\pm 1/2$ (the letter $b$), 
distributed suitably so as satisfy Eq.~\eqref{difference} consistently at every 
intermediate vertex on the original lattice. 
The grey shaded clusters in Fig.~\ref{ampdistribution1}(a) embrace the 
non-zero amplitudes only, while every yellow shaded zone is surrounded by vertices 
where amplitudes are zero. The significant observation is that, clusters of non-
zero amplitude span over a finite distance, but ultimately get decoupled from each 
other on a larger scale of length. This can be appreciated if we look at 
Fig.~\ref{ampdistribution1}(b) which is a larger version of the previous figure. 
The grey shaded clusters are distributed along the principal $X$- and $Y$-axes, 
but are {\it decoupled} from each other beyond a certain extent by the unfilled 
white boxes. The yellow clusters representing amplitude-voids are now seen to span 
larger spatial distances. A similar construction is possible for $E=-\sqrt{6}$ 
which is another solution of Eq.~\eqref{roots} for $n=1$.

\subsection{Staggered localization}
It is appearent from the above discussion that the eigenfunction corresponding to 
$E=\pm \sqrt{6}$ will be localized in the fractal space, as 
the spanning clusters of different sizes and embracing non-zero amplitudes  
ultimately get decoupled from one another. This is easily re-confirmed by studying 
the evolution of the hopping integrals under successive RSRG steps. The hopping 
integrals $t$ and $\lambda$ (zero initially, but grows later) 
remain non-zero at the first stage of RSRG (that is, $n=1$), indicating that the 
nearest neighboring sites on a one step renormalized lattice will have a non-zero 
overlap of the eigenfunctions. They start decaying for $n\geq 2$ with the decay in 
$\lambda(n)$ taking place at a much slower rate compared to $t(n)$. This indicates 
that over larger scale of length the corresponding states are ultimately 
{\it localized}, but the effect is a weak one.

This observation immediately leads to an innovative way of exactly determining a 
localized eigenstate on such a deterministic geometry. It should be appreciated 
that though it is not unnatural that most of the single particle states will be 
Anderson-localized in the absence of any translational order, nevertheless an 
exact prescription of the determination of any localized eigenvalue is not easy to 
obtain, and has not been reported so far in the literature, to  the best of our 
knowledge. We do it using the following method.

We can solve Eq.~\eqref{roots} in principle, for any $n$. For example, we have 
done it explicitly for $n=1$, $n=2$, $n=3$ and $n=4$. 
With the same set of parameters as discussed above, the roots
of Eq.~\eqref{roots} for $n=2$ are, $E=0$, $\pm \sqrt{6}$, $\pm 2.11619$, 
$\pm 0.77508$, $\pm 2.98681$. As we observe, the roots for the $n=1$ stage, viz., 
$E=0$ and $\pm \sqrt{6}$ are included in this set for $n=2$. $E=0$ corresponds
to the extended state and $E=\pm \sqrt{6}$ provide two localized states
we already know. For each of the additional roots, viz., $E= \pm 2.11619$, 
$\pm 0.77508$, $\pm 2.98681$, the hopping integrals $t$ and $\lambda$ 
remain non-zero (with considerable magnitude) up to the second stage of iteration 
($n=2$), and starts to lose their `strengths' as the renormalization progresses. 
Finally, for large `$n$' the hopping integrals become zero.

\begin{figure}[ht]
\includegraphics[clip,width=8cm]{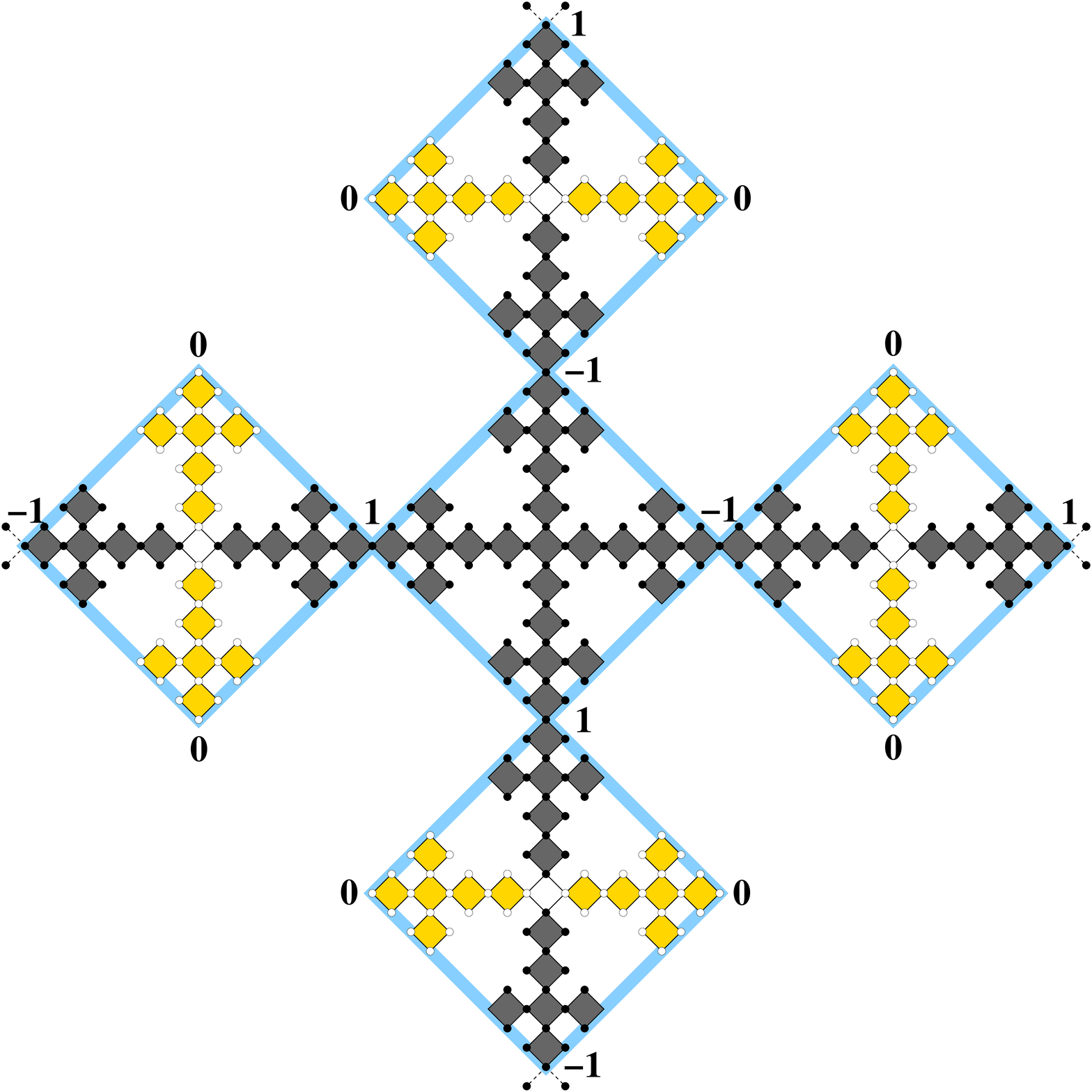}
\caption{ (Color online) Distribution of amplitudes of the wave functions at 
$E=0.77508$ (obtained by solving Eq.~\eqref{roots} for $n=2$) on a $3$rd 
generation lattice. The other parameters and symbols are same as 
FIG.~\ref{ampdistribution1}. The thick, highlighting blue lines represent a two-step renormalized lattice.}
\label{ampdistribution2}
\end{figure}

The above observation implies that, using a subset of energy values extracted 
at the stage $n=2$ ($E = \pm 2.11619$,$\pm 0.77508$ and $\pm 2.98681$), we can 
work out eigenfunctions which will span bigger clusters of lattice points 
on the original lattice compared to those obtained from $n=1$. The states will 
appear to be `extended' when viewed on a finite diamond-Vicsek fractal at the 
second generation, but will eventually be localized on a lattice in the 
thermodynamic limit. In Fig.~\ref{ampdistribution2}
we show the distribution of amplitudes for $E=0.77508$, a value that is 
obtained from Eq.~\eqref{roots} for $n=2$. The enlargement in the cluster-size 
having non-zero values of the amplitude in comparison to the $n=1$ case 
(Fig.~\ref{ampdistribution1}(b)) is obvious. The spanning clusters finally 
get decoupled from each other, just as it was for the $n=1$ case. But, now this 
decoupling occurs at a larger length scale.

It is now easy to foresee what is going to happen for $n=3$, $4$ and beyond. For 
any $n=\ell$ we will be getting roots of Eq.~\eqref{roots}, subsets of which are  
solutions of Eq.~\eqref{roots} for  $n = 1$, $2$, $\hdots$ , $\ell-1$. For these 
subsets, the decay in the hopping integrals will begin at $n > 1$, $n > 2$, 
$\hdots$, $n > \ell-1$. For the roots in addition to these, the hopping integrals 
lose their strengths and finally decay, from $n > \ell$. Thus, the latter 
eigenvalues will correspond to localized eigenstates, the localization being {\it 
delayed} (staggered) in space with localization lengths much larger than the 
previous ones. When mapped back on to the original lattice, the amplitudes for 
these additional roots will be found to span clusters of increasing size. 
The exact size of the spanning clusters will be determined  by the value of $n$. 

\begin{figure}[ht]
\centering 
\includegraphics[clip,width=6cm,angle=-90]{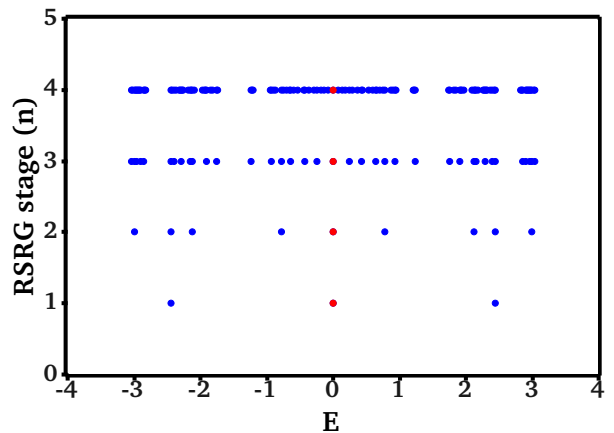}
\caption{ (Color online) Distribution of energy eigenvalues $E$ obtained from the 
Eq.~\eqref{roots} for different RSRG stage $n$. The central dot (red one) 
at $E=0$ represents the eigenvalue for the extended eigenstate.}
\label{stagloc}
\end{figure}

The roots of Eq.~\eqref{roots} are found to cluster around the value $E=0$ 
symmetrically, and tend to densely fill the neighborhood of $E=0$, at which the 
single extended eigenstate determined so far resides. The clustering of the 
eigenvalues is shown in Fig.~\ref{stagloc}. 
This dense filling of the eigenvalue spectrum around the center is also reflected 
in the apparent continuum observed in the density of states (Fig.~\ref{ldos1}(b)).

\section{Spectral properties with non-zero magnetic field}
\subsection{The energy eigenvalue spectrum}
\begin{figure}[ht]
\includegraphics[clip,width=6cm,angle=-90]{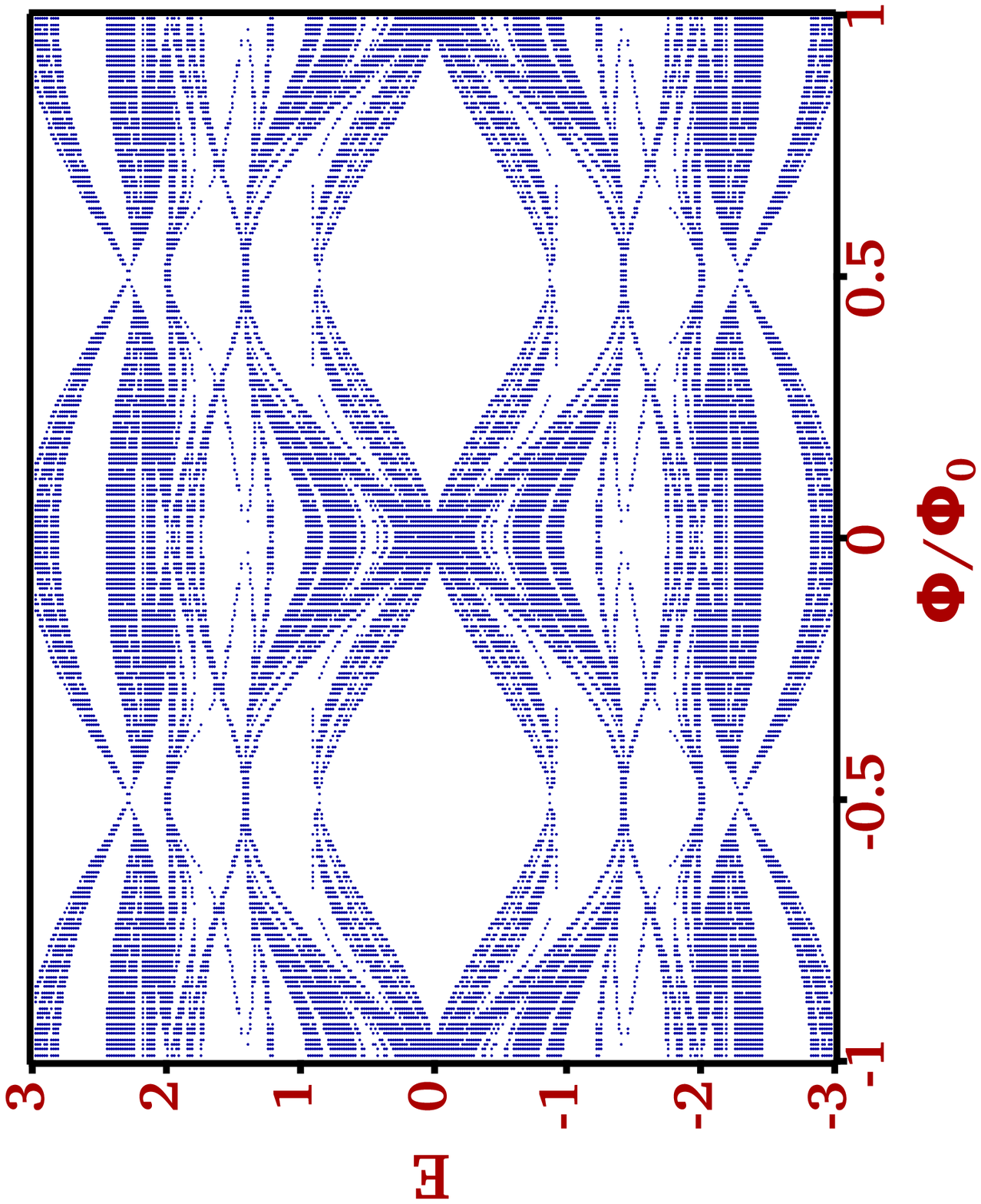}
\caption{ (Color online) Energy eigenvalue spectrum of an infinite diamond-
Vicsek fractal as a function the magnetic flux $\Phi$. We have chosen 
$\epsilon_{A}=\epsilon_{B}=0$, $t=1$ and $\lambda=0$.}
\label{engspec}
\end{figure}

We have obtained the energy eigenvalue distribution (Fig.~\ref{engspec}) 
as a function of the magnetic flux $\Phi$ enclosed in each basic plaquette for an 
infinite size diamond-Vicsek fractal. We have examined the formation of the bands 
and the gaps with the variation of magnetic flux $\Phi$. To obtain the energy 
spectrum, we have calculated the local density of states (LDOS) at both `$A$' 
and `$B$' sites by fixing the value of the energy $E$ and varying the magnetic 
flux $\Phi$ from $-1$ to $1$, repeated the above process for different values 
of energy $E$, and picked up those values of energy $E$ and magnetic flux $\Phi$  
for which we get a non-zero LDOS either at an $A$-site or a $B$-site. Thus 
Fig.~\ref{engspec} is representative of an infinite lattice. 

In Fig.~\ref{engspec}, we can clearly observe the formation of multiple bands and 
gaps, and how a variation of magnetic flux $\Phi$ leads to band overlapping. The 
band crossing is maximum at the center (around $\Phi=0$) and the density of 
allowed energy eigenvalues is large is this area. As we shift form $\Phi=0$ on 
either side, there is thinning of allowed energy eigenvalues. And finally at 
$\Phi=\Phi_{0}/2$, only four energy eigenvalues are allowed indicating an extreme 
localization of the electronic states which is discussed in details in the 
next subsection. Fig.~\ref{engspec} corroborates this last observation, though 
due to limit of resolution of the diagram, the four eigenvalues are not clearly 
seen there. This is resolved in the next diagram (Fig.~\ref{ldos2}(b)). 

\subsection{Extreme localization of the electronic states}
\begin{figure}[ht]
\centering 
\includegraphics[clip,width=10cm,angle=-90]{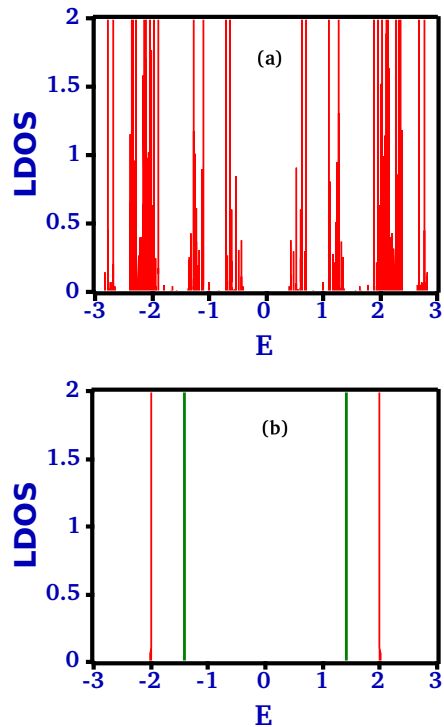}
\caption{ (Color online) (a) LDOS-$E$ plot at the bulk sites ($B$-type) of 
an infinite diamond-Vicsek network with $\Phi=\Phi_{0}/4$, and (b) is for 
$\Phi=\Phi_{0}/2$. The red lines correspond to LDOS at $B$-sites and the 
green lines correspond to LDOS at $A$-sites. We have chosen 
$\epsilon_{A}=\epsilon_{B}=0$, $t=1$ and $\lambda=0$.}
\label{ldos2}
\end{figure}

In absence of magnetic field, there was clearly a non-zero value of LDOS at the 
center of the spectrum (around $E=0$) (Fig.~\ref{ldos1}(a)). As soon as the 
magnetic field is switched on, a wide gap opens up in the LDOS spectrum around 
$E=0$ (Fig.~\ref{ldos2}(a)). 
The gap becomes wider as we increase the value of magnetic flux $\Phi$,  
finally leading to an extreme localization of electronic states (Fig.~\ref{ldos2}
(b)) at the half flux quantum (i.e., $\Phi=\Phi_{0}/2$). The four separate lines 
in Fig.~\ref{ldos2}(b) (at $E= \pm \sqrt{2}$ and at $E= \pm 2$) correspond to four 
highly degenerate localized states pinned at the `$A$'- and `$B$'-sites 
respectively. This observation is in accordance with Vidal {\it et al.}'s 
result~\cite{vidal}, where the Aharonov-Bohm caging of the 
localized orbitals under the action of an external magnetic field was discussed. 

The origin of the four localized state eigenvalues above can be easily explained 
if one appreciates that at $\Phi = \Phi_0/2$, the effective coupling between the 
sites at the vertices of an elementary rhombus, viz., 
$t^{\textit{eff}} = [2t^2/(E-\epsilon_A)] \cos(\pi\Phi/\Phi_0)$,  
becomes equal to zero. In that case one is left with only two types of atomic 
sites, decoupled from each other, and having effective on-site potential energies 
\begin{eqnarray}
\epsilon_{A}^{\textit{eff}} & = & \epsilon_A + \dfrac{2t^2}{E - \epsilon_B} \nonumber \\
\epsilon_{B}^{\textit{eff}} & = & \epsilon_B + \dfrac{4t^2}{E - \epsilon_A}
\label{extreme}
\end{eqnarray}
for the `edge' and the `bulk' sites respectively. 
With $\epsilon_A = \epsilon_B = 0$ and $t = 1$, the localized states are obtained 
by setting $E = \epsilon_{A}^{\textit{eff}}$ and 
$E = \epsilon_{B}^{\textit{eff}}$, which yield the 
values $E = \pm \sqrt{2}$ and $\pm 2$ respectively. These are the energy 
eigenvalues at which extreme localization is observed, as shown in 
Fig.~\ref{ldos2}(b).  

\section{Two terminal conductance for a finite lattice}
To get the two terminal conductance for a finite size diamond-Vicsek fractal, we 
attach the system between two semi-infinite one-dimensional ordered metallic 
leads, namely, the source and the drain. The leads, in the tight binding model, 
are described by a constant on-site potential $\epsilon_{l}$ and a nearest 
neighbor hopping integral $t_{l}$. We then successively renormalize the system to 
reduce it into an effective di-atomic system~\cite{bibhas}, consisting of two 
‘renormalized’ atoms, each having an effective on-site potential equal to $U$ and 
with an effective hopping integral $V$ between them. The transmission coefficient 
across the effective dimer is given by the well known formula~\cite{stone},
\begin{widetext}
\begin{equation}
T=\dfrac{4\sin^{2}ka}{\left[(M_{12}-M_{21})+(M_{11}-M_{22})\cos ka 
\right]^{2}+\left[(M_{11}+M_{22})\sin ka \right]^{2}}
\end{equation}
\end{widetext}
where,
$M_{11} =\dfrac{(E-U)^{2}}{Vt_{l}}
-\dfrac{V}{t_{l}},\ 
M_{12} =-\dfrac{(E-U)}{V},\ 
M_{21} =-M_{12},\ 
M_{22} =-\dfrac{t_{l}}{V}$ are the matrix elements of the transfer matrix for the 
effective di-atomic system, and $\cos ka = (E-\epsilon_{l})/2t_{l}$, `$a$' being 
the lattice constant and is taken to be equal to unity throughout the calculation.

\begin{figure}[ht]
\centering 
\includegraphics[clip,width=10cm,angle=-90]{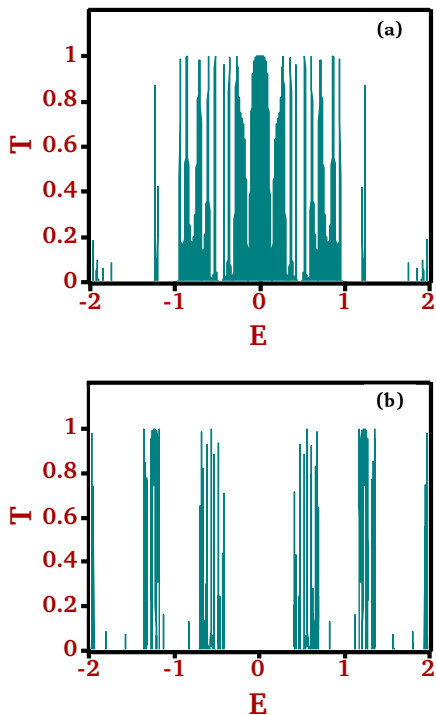}
\caption{ (Color online) Transmission characteristics for a $3$rd generation 
system. (a) is for $\Phi=0$ and (b) is for $\Phi=\Phi_{0}/4$. The other 
parameters are $\epsilon_{A}=\epsilon_{B}=0$, $t=1$ and $\lambda=0$. The 
lead parameters are $\epsilon_{l}=0$ and $t_{l}=1$.}
\label{trans}
\end{figure}

In Fig.~\ref{trans}, we have shown the two-terminal transmission characteristics 
for a 3rd generation system for different values of magnetic flux $\Phi$. For 
$\Phi=0$, the system exhibits a continuous high transmission window over a 
region at the center (Fig.~\ref{trans}(a)). 
This is due to the fact that on either side of $E=0$, the energy eigenvalues 
become quite densely packed. The corresponding eigenfunctions have localization 
lengths extending much beyond the 3rd generation 
fractal.

As we tune the magnetic flux to a non-zero value, e.g., $\Phi=\Phi_{0}/4$, 
the transmittivity of the system drastically decreases (Fig.~\ref{trans}(b)) and 
with increase in $\Phi$, the value of the transmission coefficient $T$ decreases 
more and more and finally at $\Phi=\Phi_{0}/2$, the system becomes completely 
opaque to an incoming electron. So by fixing the Fermi level of the electron to a 
particular energy, say at $E=0$, one can easily transform the system from a 
conducting one to an insulating one by tuning the external magnetic flux $\Phi$ 
suitably. 
\section{Concluding remarks}
In conclusion, we have examined the energy spectrum of a Vicsek 
geometry consisting of diamond shaped loops. The major result is that 
we have been able to identify a countable infinity of localized eigenstates
displaying a multitude of localization lengths. A prescription is given for 
an exact determination of the eigenvalues corresponding to all such states, 
a problem that is far from trivial in the case of a deterministically disordered 
system. The localized states span across the fractal space in clusters of 
increasing sizes, the size being precisely controlled by the length scale 
at which the energy eigenvalue is extracted. The onset of localization can 
be exactly predicted from the stage of RSRG. In addition, the application 
of a uniform magnetic field perpendicular to the plane of the fractal is 
found to produce gaps in the energy spectrum. A special value of the 
magnetic flux viz., $\Phi = \Phi_0/2$ is shown to lead to an extreme 
localization of the electron states as well. The results are corroborated by the 
density of states calculations, and the valuation two terminal magneto-transport.
\begin{acknowledgments}
Biplab Pal would like to thank DST-INSPIRE Program, India 
for providing financial assistance through INSPIRE Fellowship (IF110078).
Illuminating conversation with Bibhas Bhattacharyya is gratefully 
acknowledged. 
\end{acknowledgments}


\begin{thebibliography}{99}
\bibitem{anderson} P. W. Anderson, Phys. Rev. \textbf{109}, 1492 (1958).
\bibitem{abrahams} E. Abrahams, P. W. Anderson, D. C. Licciardello, and 
T. V. Ramakrishnan, Phys. Rev. Lett. \textbf{42}, 673 (1979).
\bibitem{angus} A. MacKinnon and B. Kramer, Phys. Rev. Lett. \textbf{49}, 695 
(1982).
\bibitem{thouless} D. J. Thouless, Phys. Rev. Lett. \textbf{61}, 2141 (1988).
\bibitem{schreiber} I. V. Plyushchay, R. A. R\"{o}mer, and M. Schreiber, 
Phys. Rev. B \textbf{68}, 064201 (2003).
\bibitem{alberto1} A. Rodriguez, L. J. Vasquez, K. Slevin, and R. A. R\"{o}mer,
Phys. Rev. Lett. \textbf{105}, 046403 (2010).
\bibitem{alberto2} A. Rodriguez, L. J. Vasquez, K. Slevin, and R. A. R\"{o}mer,
Phys. Rev. B \textbf{84}, 134209 (2011).
\bibitem{zilly} M. Zilly, O. Ujs\'{a}ghy, M. Woelki, and D. E. Wolf, 
Phys. Rev. B. \textbf{85}, 075110 (2012).
\bibitem{jovic} D. M. Jovi\'{c}, M. R. Beli\'{c}, and C. Denz, Phys. Rev. A 
\textbf{85}, 031801 (2012).
\bibitem{bliokh} K. Y. Bliokh, S. A. Gredeskul, P. Rajan, I. V. Shadrivov, and 
Y. S. Kivshar, Phys. Rev. B \textbf{85}, 014205 (2012).
\bibitem{echeverria} C. Echeverria-Arrondo and E. Ya. Sherman, Phys. Rev. B 
\textbf{85}, 085430 (2012).
\bibitem{sankar1} T. A. Sedrakyan, J. P. Kestner, and S. Das Sarma, Phys. Rev. A 
\textbf{84}, 053621 (2012).
\bibitem{aubry} S. Aubry and G. Andr\'{e}, Ann. Israel Phys. Soc. \textbf{3}, 133 
(1980).
\bibitem{sankar2} S. Das Sarma, S. He, and X. C. Xie, Phys. Rev. B \textbf{41}, 
5544 (1990).
\bibitem{billy} J. Billy, V. Josse, Z. Zuo, A. Bernard, B. Hambrecht, P. Lugan, 
D. Cl\'{e}ment, L. Sanchez-Palencia, P. Bouyer, and A. Aspect, Nature (London) 
\textbf{453}, 891 (2008).
\bibitem{edwards} E. E. Edwards, M. Beeler, T. Hong, and S. L. Rolston, 
Phys. Rev. Lett. \textbf{101}, 260402 (2008).
\bibitem{roati} G. Roati, C. D'Errico, L. Fallani, M. Fattori, C. Fort, M. 
Zacanti, G. Modugno, M. Modugno, and M. Inguscio, Nature (London), \textbf{453}, 
895 (2008).
\bibitem{shi} Z. Shi and A. Z. Genack, Phys. Rev. Lett. \textbf{108}, 043901 
(2012).
\bibitem{asatryan} A. A. Asatryan, L. C. Botten, M. A. Byrne, V. D. Freilikher, 
S. A. Gredeskul, I. V. Shadrivov, R. C. McPhedran, and Y. S. Kivshar, Phys. Rev. B 
\textbf{85}, 045122 (2012).
\bibitem{dunlap} D. H. Dunlap, H-L. Wu, and P. W. Phillips, Phys. Rev. Lett. 
\textbf{65}, 88 (1990); D. H. Dunlap, K. Kundu, and P. W. Phillips, Phys. Rev. B 
\textbf{40}, 10999 (1989).
\bibitem{moura1} F. A. B. F. de Moura and M. L. Lyra, Phys. Rev. Lett. 
\textbf{81}, 3735 (1998).
\bibitem{arunava1} A. Chakrabarti, S. N. Karmakar, and R. K. Moitra, 
Phys. Rev. Lett. \textbf{74}, 1403 (1995).
\bibitem{arunava2} A. Chakrabarti, S. N. Karmakar, and R. K. Moitra, 
Phys. Rev. B \textbf{50}, 13276 (1994).
\bibitem{macia1} A. S\'{a}nchez, E. Maci\'{a}, and F. Dominguez-Adame, 
Phys. Rev. B \textbf{49}, 147 (1994).
\bibitem{macia2} E. Maci\'{a}, Phys. Rev. B \textbf{60}, 10032 (1999).
\bibitem{macia3} E. Maci\'{a}, Aperiodic Structures in Condensed Matter, CRC Press (Florida) (2009).
\bibitem{arunava3} A. Chakrabarti, J. Phys.: Condens. Matt. \textbf{8}, 10951 
(1996).
\bibitem{arunava4} A. Chakrabarti, Phys. Rev. B \textbf{72}, 134207 (2005).
\bibitem{schwalm1} W. A. Schwalm and B. J. Moritz, Phys. Rev. B \textbf{71}, 
134207 (2005)
\bibitem{sil1} S. Sil, S. K. Maiti and A. Chakrabarti, Phys. Rev. Lett. 
\textbf{101}, 076803 (2008).
\bibitem{moura2} F. A. B. F. de Moura, R. A. Caetano, and M. L. Lyra, 
Phys. Rev. B \textbf{81}, 125104 (2010).
\bibitem{vicsek} T. Vicsek, Fractal Growth Phenomena (2nd Ed.), World Scientific, 
Singapore (1992).
\bibitem{rammal} R. Rammal and G. Toulouse, Phys. Rev. Lett. \textbf{49}, 1194 (1982).
\bibitem{domany} E. Domany, S. Alexander, D. Bensimon, and L. P. Kadanoff, 
Phys. Rev. B \textbf{28}, 3110 (1983).
\bibitem{tree} Y. Lin, B. Wu, and Z. Zhang, Phys. Rev. E \textbf{82}, 031140 
(2010).
\bibitem{jayanthi} C. S. Jayanthi and S. Y. Wu, Phys. Rev. B \textbf{48}, 10188 
(1993).
\bibitem{sil2} S. Sil, S. K. Maiti, and A. Chakrabarti, Phys. Rev. B \textbf{79}, 
193309 (2009).
\bibitem{aharony1} A. Aharony, Ora Entin-Wohlman, Y. Tokura, and S. Katsumoto, 
Phys. Rev. B \textbf{78}, 125328 (2008).
\bibitem{aharony2} A. Aharony, Ora Entin-Wohlman, Y. Tokura, and S. Katsumoto, 
Physica E \textbf{42}, 629 (2010).
\bibitem{vidal} J. Vidal, R. Mosseri, and B. Doucot, Phys. Rev. Lett. 
\textbf{81}, 5888 (1998). 
\bibitem{peierls} R. Peierls, Z. Phys. \textbf{80}, 763 (1933). 
\bibitem{southern} B. W. Southern, A. A. Kumar, and J. A. Ashraff, 
Phys. Rev. B \textbf{28}, 1785 (1983).
\bibitem{bibhas} A. Chakrabarti and B. Bhattacharyya, 
Phys. Rev. B \textbf{54}, R12625 (1996).
\bibitem{anirban} A. Chakraborti, B. Bhattacharyya, and A. Chakrabarti, 
Phys. Rev. B \textbf{61}, 7395 (2000).
\bibitem{stone} A. D. Stone, J. D. Joannopoulos, and D. J. Chadi, 
Phys. Rev. B \textbf{24}, 5583 (1981)
\end{thebibliography}
\end{document}